\documentclass[mathleft]{an}
\usepackage{graphicx}
\usepackage{times}
\usepackage{natbib}
\begin{document}

\Pagespan{789}{}
\Yearpublication{2006}%
\Yearsubmission{2005}%
\Month{11}%
\Volume{999}%
\Issue{88}%

\title{The diversity of light curve variations of Blazhko stars}

\author{\'A. S\'odor\inst{1}\fnmsep\thanks{Corresponding author:
  \email{sodor \_at\_ konkoly \_dot\_ hu}\newline}
}
\titlerunning{The diversity of light curve variations of Blazhko stars}
\authorrunning{\'A. S\'odor}
\institute{
Konkoly Observatory, H-1525 Budapest, PO Box 67, Hungary
}

\received{}
\accepted{}
\publonline{}


\abstract{
  The ground-based Konkoly Blazhko Survey I and II aim to collect accurate, extended, multicolour light curves of galactic field RRab stars in order to determine the incidence rate of the modulation in the sample, to study the modulation in details, to study long-term changes in the modulation properties and to find changes in the mean global physical parameters of the stars with Blazhko phase. Here the diversity of light curve variations of Blazhko stars is demonstrated through a sub-sample of the modulated RRab stars found by the first part of this survey.
}

\maketitle

\section{Introduction}

The cyclic variations in the shape of RR Lyrae light curves, the Blazhko effect, recognized by \cite{blazhko} and \cite{shapley}, is one of the most long-lasting riddle of stellar pulsation theory. Recent extended studies of these stars revealed that the phenomenon manifests itself in many different ways, which poses a great challenge against theories trying to explain the effect \citep{kovacs}.

In this paper the Konkoly Blazhko Survey I and II is introduced in brief and the diversity of modulated RRab light curves is demonstrated through a sub-sample of the Blazhko stars found by the survey.

\section{The Survey}

The ground-based Konkoly Blazhko Survey (KBS) I and II aim to collect accurate, extended, multicolour light curves of northern fundamental mode RR~Lyrae stars of the Galactic field \cite{an} in order to:

\begin{itemize}
 \item determine the incidence rate of the modulation in the sample,
 \item study the modulation in details,
 \item study long-term changes in the modulation properties,
 \item find changes in the mean global physical parameters of the stars with Blazhko phase.
\end{itemize}

The PIs of the project are Johanna Jurcsik and B\'ela Szeidl, co-workers are \'Ad\'am S\'odor, Zsombor Hurta and others.

The first part of the survey, KBS I, was initiated in January 2004 using Johnson-Cousins $BV(RI)_\mathrm{C}$ filters and a Wright CCD camera attached to the automatic 60\,cm Heyde-Zeiss telescope of Konkoly Observatory at Budapest, Hungary. The first part is now finished, covering 30 bright short-period ($P_\mathrm{puls} < 0.5$\,d) variables with about 750 nights of observations. A surprisingly high number of stars (14 out of 30, 47\%) has been found to show the Blazhko effect \citep{kbs1}.

In January 2009, the second part of the survey, KBS~II, was launched extending the sample toward longer periods ($0.55$\,d\,$> P_\mathrm{puls} > 0.6$\,d) using the 1\,m RCC telescope of Konkoly Observatory at Piszk\'es-tet\H o and also the 60\,cm Heyde-Zeiss telescope. Up to now 21 RRab stars have been observed in the frame of the ongoing KBS~II project.

\section{The diversity of the modulation}

\subsection{The sample}

We demonstrate the diversity of light curve variations of Blazhko stars through a sub-sample of the modulated RRab stars found by the KBS I. Four weakly modulated (SS~Cnc, DM~Cyg, RR~Gem, BR~Tau) and four strongly modulated (XY~And, AQ~Lyr, MW~Lyr, UZ~Vir) stars have been selected. Our results on the analysis of the light curve variations of several of these stars have already been published (SS~Cnc -- \hbox{\citealt{ssc}}, DM~Cyg -- \citealt{dmc}, RR~Gem -- \citealt{rrg}, MW~Lyr -- \citealt{mw1}). All the selected variables were extensively observed. The number of data points in $V$ band are between 1400 and 5800 for these objects. The data cover all the pulsation phases in each Blazhko phase for each object. These Blazhko stars have one dominant modulation period, however, some of them have one or more secondary modulations also. We have found further Blazhko stars with multiperiodic or irregular modulation in our survey (RY~Com -- \citealt{kbs1}, V759~Cyg -- \citealt{kbs1}, CZ~Lac -- \citealt{czl0}, AQ~Lyr -- \citealt{kbs1}, UZ~UMa -- \citealt{uzu}), which adds even more colour to our present topic. We only mention the existence of the multiperiodic modulation here without giving details and we are going to analyze these objects later in forthcoming papers.

\begin{figure*}
\includegraphics[width=170mm,height=200mm]{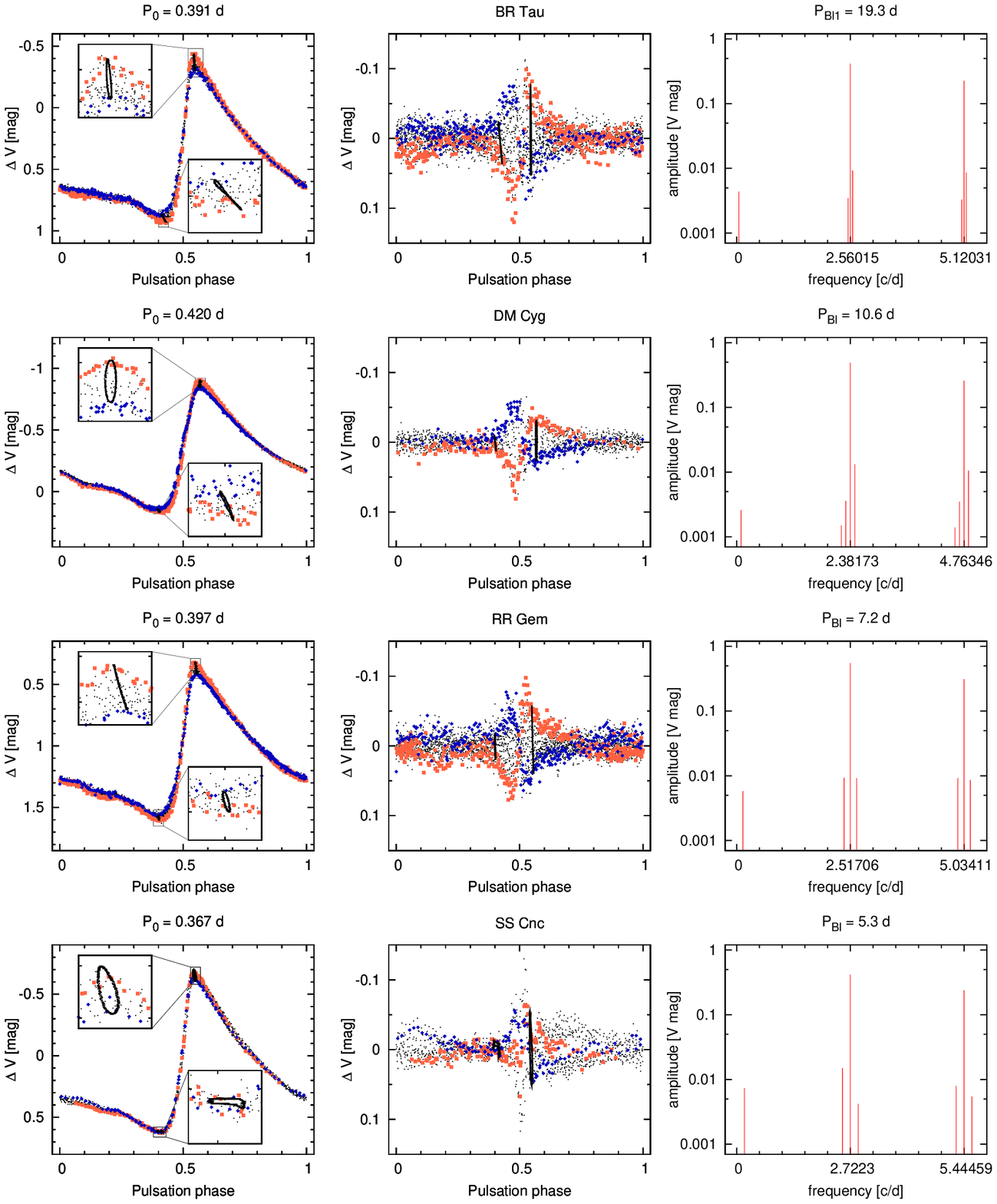}
\caption{Weakly modulated sample Blazhko stars from KBS I. The left-hand panels display the $V$ light curves folded with the pulsation period. The middle panels show the light curves after the subtraction of the mean pulsation curves. In the left-hand and middle panels the highest and lowest pulsation amplitude Blazhko phases are highlighted. The loops, which the maximum and minimum lights go along during the modulation, are also drawn. The right-hand panels plot the schematic Fourier spectra of the light curve solutions in the vicinity of the pulsation frequency ($f_0$) and its first harmonic ($2f_0$).}
\label{fig:kicsi}
\end{figure*}

\begin{figure*}
\includegraphics[width=170mm,height=200mm]{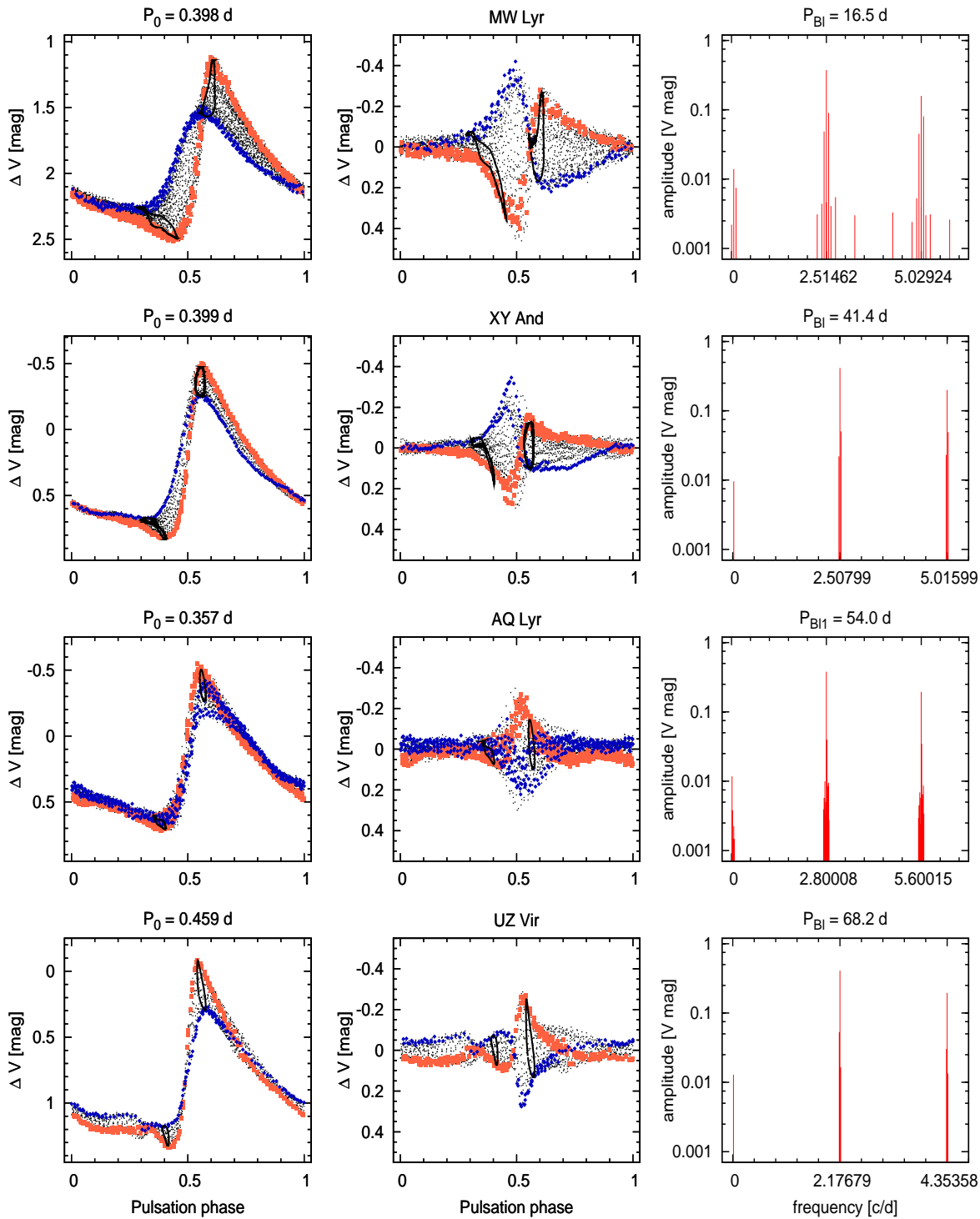}
\caption{Strongly modulated sample Blazhko stars from KBS I. The left-hand panels display the $V$ light curves folded with the pulsation period. The middle panels show the light curves after the subtraction of the mean pulsation curves. In the left-hand and middle panels the highest and lowest pulsation amplitude Blazhko phases are highlighted. The loops, which the maximum and minimum lights go along during the modulation, are also drawn. The right-hand panels plot the schematic Fourier spectra of the light curve solutions in the vicinity of the pulsation frequency ($f_0$) and its first harmonic ($2f_0$).}
\label{fig:nagy}
\end{figure*}

The weakly and strongly modulated sample objects are shown in Figs.~\ref{fig:kicsi}~and~\ref{fig:nagy}, respectively. The left-hand panels display the $V$ light curves folded with the pulsation period. The middle panels show the light curves after the subtraction of the mean pulsation curves. In the left-hand and middle panels the highest and lowest pulsation amplitude Blazhko phases are highlighted. The loops, which the maximum and minimum lights go along during the modulation, are also drawn. These loops are calculated from synthetic light curves taking into account the dominant modulation period only. The right-hand panels plot the schematic Fourier spectra of the light curve solutions in the vicinity of the pulsation frequency ($f_0$) and its first harmonic ($2f_0$). The peaks in these panels represent the frequencies and amplitudes of the components of the light curve solutions.

\subsection{Discussion}

\subsubsection*{Phase modulation}

Examining the weakly modulated stars we find that only slight, if any, phase modulation occurs. BR~Tau and RR~Gem has fix points on their rising branches, and their maximum brightness phases are also marginally modulated only. The rising branches of DM~Cyg and SS~Cnc show weak but definite phase modulation.

At the same time, all the strongly modulated Blazhko stars show definite phase modulation. A fix point can be seen on the lower part of the rising branch of UZ~Vir, while the maximum and minimum light phases are modulated in all the four objects.

\subsubsection*{Loops of maximum and minimum lights}

It is an interesting observation that the direction of the loop the maximum light goes along during the modulation is counterclockwise in each but one case, for non-degenerated loops. The only exception is UZ~Vir. In accordance with this observation, the higher frequency modulation peaks (right side, $kf_0+f_\mathrm{m}$) usually have higher amplitudes than the lower frequency ones (left side, $kf_0-f_\mathrm{m}$). On the mathematical background see details at \cite{bela}. The minimum light travels an opposite direction than the maximum light does in each cases.

\subsubsection*{Complex modulation}

The modulation of AQ~Lyr is rather complex. The dominant modulation period of this star is 54\,d but the light variation, which have been followed in two consecutive observing seasons, can be described with sufficient accuracy only when four or more modulations are taken into account. Even so, variations in the residual exceeding the accuracy of our measurements remain. BR~Tau also shows a similarly complex modulation, however, it is less conspicuous as the overall strength of its modulation is much weaker. The residual light curve of MW~Lyr, after the removal of the best solution, shows also a kind of stochastic behaviour \citep{mw1}. There are signs of secondary modulation periods also at other sample stars (MW Lyr~\citealt{mw1}, RR~Gem \citealt{rrg}) indicating that multiperiodic modulation is much more the rule than the exception in Blazhko stars.

\acknowledgements

The financial support of the Hungarian OTKA grant T-068626 is acknowledged.

\end{document}